\documentclass[twocolumn, prl, superscriptaddress, showpacs]{revtex4}
\usepackage{amsmath,amsthm,amsfonts,amssymb}
\usepackage{graphicx}
\usepackage{color}
\usepackage{natbib}
\usepackage{mciteplus}

\renewcommand{\vec}[1]{{\mathbf #1}}        
\newcommand{\ten}[1]{{\mathbf{#1}}}
\newcommand{\half}{\textstyle{\frac{1}{2}}}
\newcommand{\Id}{\ten{I}}

\begin{document}
 
\title{A non-monotonic constitutive model is not necessary to obtain
  shear banding phenomena in entangled polymer solutions}

\author{J.~M.~Adams} \affiliation{ Cavendish Laboratory, University of
  Cambridge, JJ Thomson Avenue, Cambridge, CB3 0HE, U.K.}

\author{P.~D.~Olmsted}
\affiliation{ School of Physics \& Astronomy, University of Leeds,
  Leeds, LS2 9JT, U.K.} 
\date{\today}
\begin{abstract}
  In 1975 Doi and Edwards predicted that entangled polymer melts and
  solutions can have a constitutive instability, signified by a
  decreasing stress for shear rates greater than the inverse of the
  reptation time.  Experiments did not support this, and more
  sophisticated theories incorporated Marrucci's idea (1996) of
  removing constraints by advection; this produced a monotonically
  increasing stress and thus stable constitutive behavior. Recent
  experiments have suggested that entangled polymer solutions may
  possess a constitutive instability after all, and have led some
  workers to question the validity of existing constitutive models.
  In this Letter we use a simple modern constitutive model for
  entangled polymers, the Rolie-Poly model with an added solvent
  viscosity, and show that (1) instability and shear banding is
  captured within this simple class of models; (2) shear banding
  phenomena is observable for \textit{weakly stable} fluids in flow
  geometries that impose a sufficiently inhomogeneous total shear
  stress; (3) transient phenomena can possess inhomogeneities that
  resemble shear banding, even for weakly stable fluids. Many of these
  results are model-independent.
\end{abstract}
\maketitle
Much of the rheology of entangled polymers solutions and melts is
captured by the molecular theory of Doi and Edwards
(DE) \cite{doiedwards}, who argued that polymers relax by curvilinear
diffusion (reptation) within a tube of the surrounding polymers.  The
DE model has a local maximum in the constitutive relation (the total
shear stress as a function of shear rate for homogeneous flows). The
resulting non-monotonic relation (\textit{e.g.} the dashed curves in
Fig.~\ref{fig:flowcurves}) leads to an instability that for many years
was not observed in experiments \cite{stratton1966dnn,
  *menezes1982nrb, *hieber1989sci, *pattamaprom2001cre}, but
nonetheless attracted attention
\cite{doi1979dcp,mcleish86,*mcleish87}.  This stress maximum is
predicted to be less pronounced or absent if the convected constraint
release (CCR) of entanglements due to flow
\cite{Marr96b,*mead98,*IannirubertoM00,likhtmangraham03,MilnerML01} is
incorporated.  A sudden release of a constraint can relax both the
orientation and conformation of a stretched polymer, which increases
the stress and, for sufficiently frequent events, eliminates the
instability.  The CCR mechanism also leads to neutron scattering
predictions that agree with experiment \cite{bent2003nmp}.  Similar
physics applies to solutions of breakable wormlike micelles, in which
the instability is well documented experimentally and leads to
\textit{shear banding}, in which a fast-flowing oriented state
coexists with a more disordered and viscous state along a stress
plateau \cite{SCM93,*catesfielding06}. There, CCR is less pronounced
because of breakage and fails to ameliorate a constitutive instability
\cite{MilnerML01}.

Recently, Wang, Hu and co-workers studied entangled solutions of a
high molecular weight (HMW) polymer in its own
oligomer\cite{tapadiawang03,tapadiawang06,boukany07,wang06}, or DNA
solutions \cite{WangMM2008}, finding a number of results that may be
consistent with instability and shear banding after all. In controlled
shear rate mode a weakly increasing stress plateau of three decades in
shear rate was found, whereas in controlled shear stress mode the
sheared solution experienced a jump in the shear rate, together with
spatially inhomogeneous birefringence \cite{tapadiawang03}. Local
velocimetry revealed spatially inhomogeneous velocity profiles in both
the transient and the steady state \cite{tapadiawang06} regimes, while
large amplitude oscillatory shear flow (LAOS) experiments showed an
inhomogeneous banding-like shear rate profile at finite frequencies
\cite{tapadiaravin06}.  Similar behavior was observed in a sliding
plate shear cell in monodisperse solutions \cite{boukany07}.
Relaxation after a step strain induced a highly inhomogeneous velocity
field with \textit{negative} local shear rates \cite{wang06}. Hu
\textit{et al.}  found similar inhomogeneous flow behavior and
possible signatures of shear banding in polymer solutions, and
wormlike micelle solutions at concentrations where severe shear
thinning, but not banding, might be expected \cite{hu07,*hu08}.

Wang \textit{et al.} could not reconcile their results with existing
theory, and proposed that the instability is a yield like effect due
to an unbalanced ``entropic retraction force'' \cite{wang07,
  *wangboukany07}.  Here we show that much of the phenomenology of
these experiments is consistent with the predictions of tube models
with CCR, perhaps as anticipated in the original theory
\cite{doi1979dcp}, without introducing new physics.

\begin{figure}[!htb]
\begin{center}
\includegraphics[width = 0.48\textwidth]{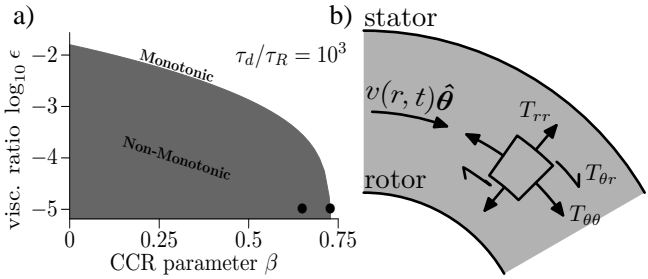}
\end{center}
\caption{a) Parameters $(\beta, \epsilon)$ where the constitutive
  curve of the Rolie-Poly model is non-monotonic (shaded). b) Section
  through Couette rheometer showing the flow field and total stress
  components, $\ten{T}$.}
\label{fig:model}
\end{figure}

\textit{Model---}We approximate the total stress $\ten{T}$ as
separating into fast Newtonian (or solvent) degrees of freedom and a
slow viscoelastic component $\ten{\Sigma}$ (HMW polymer):
\begin{equation}
\ten{T} = -p \Id + 2 \eta \ten{D} + G \ten{\Sigma},
\label{eqn:totalstress}
\end{equation}
where $\Id$ is the identity tensor, $\ten{D} = \half\left[\nabla
  \vec{v} + (\nabla \vec{v})^T\right]$, $p$ is the isotropic pressure
determined by incompressibility ($\nabla \cdot \vec{v} = 0$), and
$\eta$ is the solvent viscosity, for which we use the dimensionless
quantity $\epsilon = \eta/(G\tau_d)$. Here, $\tau_d$ is the reptation
time. We are interested in the creeping flow (low Reynolds number)
regime, in which $\nabla\cdot \ten{T}=0$.

For the dynamics of $\ten{\Sigma}$ we use the Rolie-Poly (RP) model
\cite{likhtmangraham03}, a simplified tube model that incorporates CCR
\cite{MilnerML01}, where $\ten{\Sigma}(\vec{r},t)$ obeys
\begin{multline}
  (\partial_t + \vec{v}\cdot \nabla) \ten{\Sigma}+\left(\ten{\nabla
      v}\right) \cdot \ten{\Sigma} + \ten{\Sigma}\cdot
  (\ten{\nabla v})^T + \frac{1}{\tau_d}\ten{\Sigma} = \label{eqn:RPmodel}\\
  2 \ten{D} - \frac{2}{\tau_R}(1-A) \left[\Id + \ten{\Sigma}(1+\beta
    A)\right]+\mathcal{D} \nabla^2 \ten{\Sigma},
\end{multline}
$A = ( 1+\textrm{tr} \ten{\Sigma}/3)^{-1/2}$ and the Rouse time
$\tau_R$ governs chain stretch. The stress ``diffusion'' term
${\mathcal D} \nabla^2 \ten{\Sigma}$ describes the response to an
inhomogeneous viscoelastic stress; while not in the original RP model,
it can arise due to diffusion or finite persistence length
\cite{olmsted99a,elkareh89,adams08}.  We specify Neumann boundary
conditions ($\nabla \ten{\Sigma}=0$) \cite{bhavearmstrong91,adams08}.
From experimental values of the plateau modulus $G\sim 6 \times 10^2
\textrm{Pa}$, reptation time $\tau_d \sim 20\,\textrm{s}$, and solvent
viscosity $\eta \sim 1\,\textrm{Pa s}$ \cite{tapadiawang03}, we use
$\epsilon= 10^{-5}$. Here we use $\tau_d/\tau_R \sim 10^3$, which is
consistent with the length of the stress-shear rate plateau reported
in \cite{tapadiawang03}. The parameter $\beta$ controls the efficiency
of CCR, and is difficult find a precise value of experimentally.
Ref.~\cite{likhtmangraham03} chose $\beta=1$ to fit steady state data
in polymer melts, and used multiple modes with $\beta =0.5$ to fit
experimental transient data. Here we tune between two qualitatively
differrent types of constitutive curve; either a non-monotonic
($0.65$) or a monotonic ($0.728$) constitutive curve with a broad
plateau (Figs.~\ref{fig:model},\ref{fig:flowcurves}).

Eq.~(\ref{eqn:RPmodel}) was solved in one spatial dimension using the
Crank-Nicolson algorithm \cite{olmsted99a}, for unidirectional Couette
flow $v(r,t)\widehat{\boldsymbol{\theta}}$ between cylinders of radii
$R_1$ and $R_2$ parameterized by $q\equiv \ln R_2/R_1$. In this
geometry the total shear stress $T_{r\theta}\sim1/r^2$, so that the
stress difference across the flow cell is $\Delta\ln T_{r\theta}=2 q$.
Cone and plate flow with cone angles of $\theta = (4 ^\circ ,
1^{\circ})$ has been reported \cite{tapadiawang06,hu07}, so we use
consistent values of stress difference corresponding to $q
\simeq\Delta R/R= (2 \times 10^{-3}, 2 \times 10^{-4})$
\cite{adams08}. Stresses are measured in units of $G$, shear rates in
units of $\tau_d^{-1}$, and velocities in units of $ q R_1/\tau
\approx \Delta R/\tau_d $ for small $q$. To plot numerical data we use
$\Gamma$, the dimensionless specific torque (per height per radian) on
the inner cylinder. The diffusion constant used was ${\mathcal D}
\tau_d/(R_1 q)^2 = 4\times 10^{-4}$.

\textit{Flow Curves---}To calculate the steady state flow curves a
step shear rate was applied from rest and evolved for $500 \tau_d$
with time step $10^{-5} \tau_d$, after which subsequent shear
rate steps and time evolutions were applied to scan up and down in
shear rate (Fig.~\ref{fig:flowcurves}).  For non-monotonic
constitutive curves ($\beta=0.65$) shear banding always occurs, with
hysteresis and a stress ``plateau''.  For the monotonic case
($\beta=0.728$) shear banding could be inferred in the more highly
curved geometry with the larger stress difference (larger $q$), since
the flow curve no longer follows the constitutive curve; but
\textit{without} hysteresis. Crudely, a monotonic flow curve exhibits
banding-like flows when most of the shear rate in the gap occurs over
a small range of stresses, \textit{i.e.}  the slope of the plateau
must be much smaller than the apparent slope specified by the flow
geometry:
\begin{equation}
  \left.\frac{d\Gamma}{d\dot{\gamma}}\right|_{\textrm{C.C.}}\ll
\left.\frac{\Gamma(R_1)-\Gamma(R_2)}{\Delta
    \dot{\gamma}}\right|_{\textrm{g}}
\sim e^q-1, 
\label{eq:condition}
\end{equation}
where ``C.C.'' denotes the flat portion of the constitutive curve
(dashed in Fig.~\ref{fig:flowcurves}) and ``g'' refers to the range of
torques and shear rates specified by the flow geometry.

\begin{figure}[!htb] 
  \begin{center}
    \includegraphics[width = 0.48\textwidth]{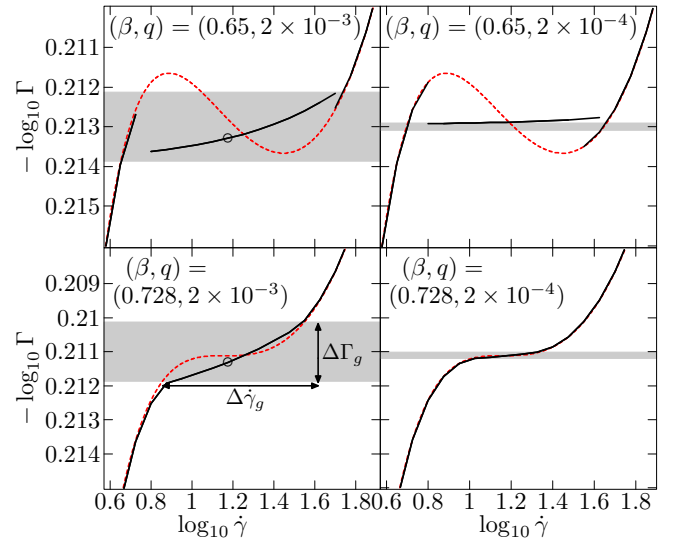}
  \end{center}
  \caption{Flow curves (solid black) and constitutive curves (dashed
    red) for a range of stress gradients $q$ and CCR values $\beta$.
    The shaded area shows the size of the torque difference $\Delta
    \log_{10} \Gamma = \log_{10} e^{2 q}$. Circles indicate
    applied shear rates at which the transient state
    responses are shown in Fig.~\ref{fig:vprofs}.}
  \label{fig:flowcurves}
\end{figure}

\begin{figure}[!htb] 
  \begin{center}
    \includegraphics[width = 0.48\textwidth]{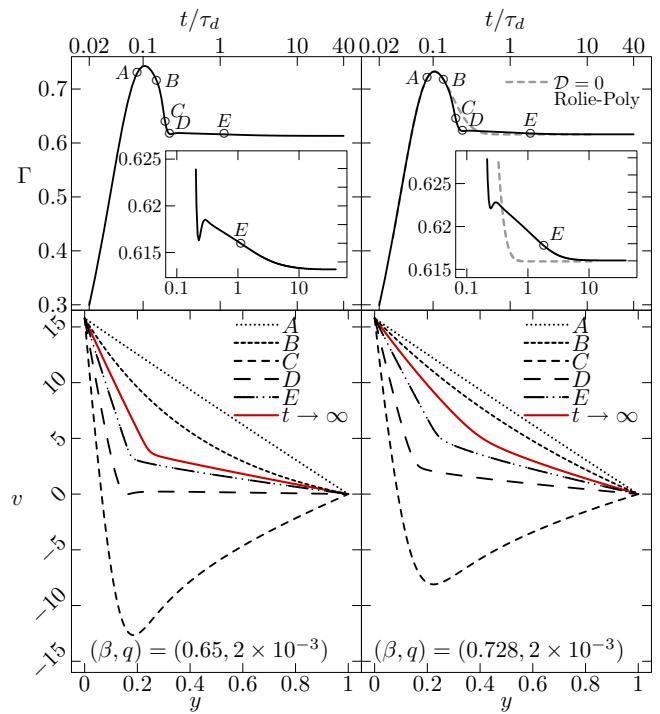}
  \end{center}
  \caption{Velocity as a function of position $y=q^{-1}\ln (r/R_1)$
    for different $(\beta, q)$, at shear rate $\log_{10}\dot{\gamma} =
    1.2$ ($\circ$ in Fig.~\ref{fig:flowcurves}). The solid (red) lines
    indicate the steady state profile, and dashed lines are transient
    profiles at the times shown. The transient torque response in the
    spatially uniform model is shown for the monotonic case
    ($\beta=0.728$) by a dashed line }
  \label{fig:vprofs}
\end{figure}

The steady state velocity profiles are shown in Fig.~\ref{fig:vprofs}
as solid (red) lines.  The non-monotonic flow curves $(\beta=0.65)$
lead to a pronounced kink in the velocity profile, a signature of
shear banding. The monotonic case does not shear band in the flatter
geometry (small $q$), but for a more curved geometry (larger $q$) more
shear rates are accessible and the resulting smooth velocity profile
could easily be interpreted as banding \cite{tapadiawang06}; certainly
the constitutive curve is not followed (Fig.~\ref{fig:flowcurves}).
Similar smooth profiles were reported in \cite{tapadiawang06,hu07}, in
a flow geometry with $q\simeq0.004-0.02$. A slightly increasing stress
plateau over several decades in shear rates (as in
\cite{tapadiawang03}) would thus lead to apparently banding
(inhomogenous) flow in geometries with very low stress gradients; but
a linear steady state profile is found if $q$ is sufficiently small
(Eq.~\ref{eq:condition}).

\textit{Startup Transients---}Transients were studied by evolving from
rest using a time step of $10^{-5}\tau_d$ (Fig.~\ref{fig:vprofs}). In
all cases shown here strongly inhomogeneous flow develops after the
stress overshoot, leading to a sharply banded transient state, with a
\textit{negative} velocity and shear rate in the less viscous band. In
the monotonic case the velocity profile eventually smooths out. For a
narrower stress plateau (\textit{e.g.}, $\beta=0.3$, not shown) the
overshoot has a less pronounced kink and typically a positive shear
rate. We have found inhomogeneous transients with negative velocities
with stress differences corresponding to a cone angle
$\theta=0.003^\circ$ in startup runs, but for $\theta = 0.001^\circ$
the amplitude of the inhomogeneous flow is reduced, and the velocity
no longer negative, whilst for $\theta = 0^{\circ}$ the flow remains
homogeneous. With perturbed initial conditions then the inhomogeneous
transient behaviour returns. The transient for a monotonic model
$(\beta=0.728)$ in which spatial gradients are artificially prohibited
exhibits a slower decrease after the stress overshoot than in the
spatially resolved model (Fig.~\ref{fig:vprofs}); hence,
inhomogeneities are important when using transient data to help
differentiate candidate constitutive models
\cite{doiedwards,WangMM2007}.

\begin{figure}[!htb] 
  \begin{center}
    \includegraphics[width = 0.48\textwidth]{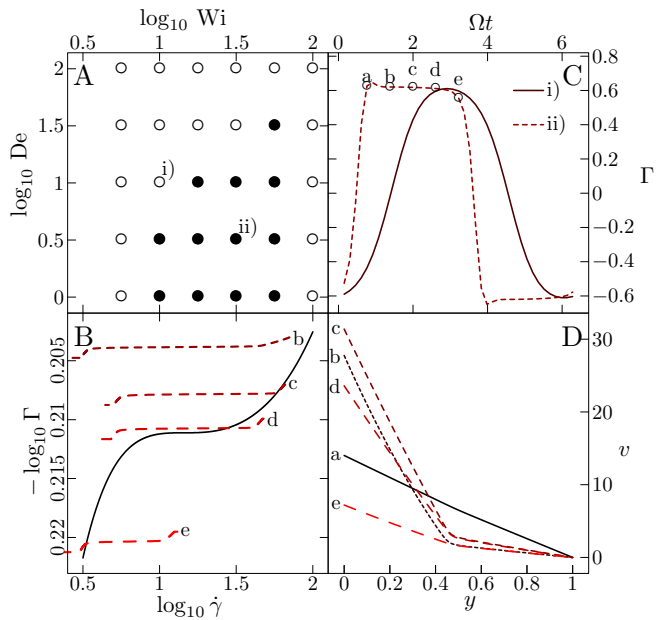}
  \end{center}
  \caption{(A) Pipkin diagram ($\mathit{Wi}$ vs $\mathit{De}$) for
    $\beta = 0.728$ and $q = 2 \times 10^{-3}$, showing regions with
    homogeneous ($\circ$) and inhomogeneous profiles ($\bullet$). (B)
    Parametric shear rate--torque profiles 
    $( \dot{\gamma}(y),\Gamma(y))$ (dashed: c,d) overlaying the
    constitutive curve (solid). (C) Torque evolution for
    different $\mathit{De}$ noted on (A). (D) Velocity
    profiles for ii); profiles (a-e) are at different times in the
    cycle for $\textit{De}\simeq3$ (ii) ($\circ$ in C). }
  \label{fig:LAOS}
\end{figure}

\textit{Large Amplitude Oscillatory Shear (LAOS)---}
A sinusoidal spatially-averaged shear rate was applied with
frequency $\Omega$ and maximum shear rate $\dot{\gamma}_m$, and
evolved from rest (zero stress) until any initial transients had
decayed.
We characterize the dynamics by the Weissenberg number $\mathit{Wi} =
\dot{\gamma}_m \tau_d$ and the Deborah number $\mathit{De} = \Omega
\tau_d$.  For low $\mathit{De}$ (frequency) we expect to recover some
features of the steady state behavior, such as transient banding for
$\mathit{Wi}$ roughly within the non-monotonic part of the flow curve;
while higher frequencies (high $\mathit{De}$) should produced sharper
profiles similar to the transient behavior in Fig.~\ref{fig:vprofs},
since the system cannot relax before flow reversal.  At the highest
frequencies we expect the reversing dynamics to be too fast to allow
an inhomogenous state. 

Fig.~\ref{fig:LAOS}A shows this behavior on a ``Pipkin diagram'' of
$\mathit{Wi}$ vs. $\mathit{De}$, for a monotonic flow curve
($\beta=0.728$) in a slightly curved geometry.  The inhomogeneous
profiles in the banding regime (Fig.~\ref{fig:LAOS}D) can be
represented parametrically in terms of shear rate and torque, $(
\dot{\gamma}(y),\Gamma(y))$ (Fig.~\ref{fig:LAOS}B).  At the high
stress regions of the cycle a portion of the sample enters the high
shear rate band as reported experimentally \cite{tapadiaravin06}. In
these calculations the position $y_{\ast}$ of the interface at a given
strain $\dot{\gamma}_0/\Omega = 3$ varied with $\mathit{De}$ as
$y_{\ast}\sim (\mathit{De})^{\alpha}$, where $\alpha\sim0.4-0.6$,
unlike the fixed position reported in Ref.~\cite{tapadiaravin06}. We
suspect that these experiments did not attain steady state.  The
torque overshoot is typical of polymer solutions, and resembles that
found in \cite{hu07} (Fig.~\ref{fig:LAOS}).  At low frequencies the
system has time to find a selected stress, which remains constant for
part of the cycle while the shear band grows into the cell.  At high
frequencies the fluid cannot relax or shear band, which leads to a
sinusoidal response and a nearly affine spatial profile.

\begin{figure}[!htb] 
  \begin{center}
    \includegraphics[width = 0.48\textwidth]{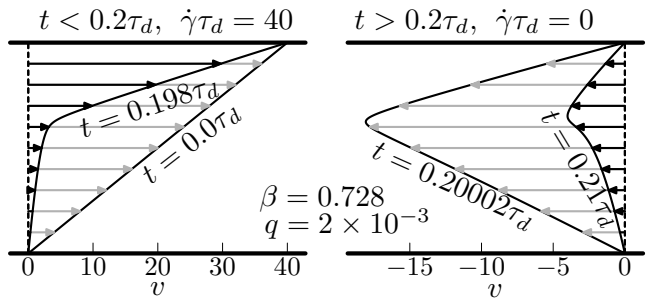}
  \end{center}
  \caption{Relaxation of a step strain of $\gamma=8$, applied using
    $\dot{\gamma} \tau_d= 40$ for $t=0.2\tau_d$.  (Left) Velocity before
    the shear rate stops and (right) snap shots of recoil velocities.}
  \label{fig:stepstrain}
\end{figure}
\textit{Step Strain---}Some experiments on the relaxation after a step
strain found a strong inhomogeneous recoil that developed a negative
velocity gradient \cite{wang06}. We illustrate this with a monotonic
constitutive curve ($\beta = 0.728$), Fig.~\ref{fig:stepstrain}. As in
Fig.~5 of \cite{wang06}, an inhomogeneous velocity profile develops
and the velocity becomes negative as the system recoils from the
applied shear.  Experimentally, the total displacement after recoil is
of the order of a tenth of the gap size which is comparable with that
observed here. Thus, inhomogeneities are important when using step
strain data to help discriminate among candidate constitutive models,
as when using the DE damping function \cite{doiedwards,WangMM2007}.

\textit{Summary---}We have shown that behavior reminiscent of shear
banding, as reported recently, can be reproduced using the Rolie-Poly
model supplemented by a term to accommodate spatial gradients.  The RP
model contains an unknown parameter $\beta$, which controls the
efficacy of convected constraint release.  Even for $\beta$ large
enough to yield a stable (monotonic) constitutive curve, shear banding
signatures can appear if the ``stress plateau'' is flat enough: (1) a
geometry with a high stress gradient can induce a flow profile that
could be mistaken for banding; (2) sharp banding-like profiles can appear
in start-up transients even though the steady state is  non-banded;
(3) LAOS can trap these sharp transient profiles; and (4) relaxation
after a  large step strain can be very inhomogeneous,
sometimes with a negative shear rate recoil. Several recent
experiments, particularly on  polydisperse polymer solutions, may
fall into this category \cite{hu07}. A wide plateau is believed to
accompany very highly entangled systems \cite{MilnerML01}, and the
larger number of relaxation times are  likely to render
polydisperse systems intrinsically more stable than monodisperse
systems \cite{doi1979dcp}, as was noted in recent experiments
\cite{hu07}. Our results are not specific to the RP model; Zhou
\textit{et al.}  recently studied a different two-fluid model of shear
banding (with a non-monotonic constitutive relation), and found
qualitative results similar to some of ours \cite{cook08}.

We thank S.-Q. Wang, R. Graham, T. McLeish, O. Radulescu, and S.
Fielding for lively discussions. This work was supported by the Royal
Commission of 1851.
\bibliography{../../banding_bib/SQW,../../banding_bib/malkus,../../banding_bib/LCtheory,../../banding_bib/sriram,../../banding_bib/worms3,../../banding_bib/rheofolks,../../banding_bib/master,../../banding_bib/worms,../../banding_bib/berrportdecruppe,../../banding_bib/callaghan,../../banding_bib/rheochaosPRL,../../banding_bib/bord04,../../banding_bib/shear05,../../banding_bib/articles,../../banding_bib/books,../LAOS}

\ifx\mcitethebibliography\mciteundefinedmacro
\PackageError{apsrevM.bst}{mciteplus.sty has not been loaded}
{This bibstyle requires the use of the mciteplus package.}\fi

\end{document}